\documentstyle[prd,aps,preprint,epsfig,floats]{revtex}
\tightenlines

\def\dsodt{ds_{23}\over dt}
\def\dstdt{ds_{13}\over dt}
\def\dsthdt{ds_{12}\over dt}

\input epsf.tex
\def\DESepsf(#1 width #2){\epsfxsize=#2 \epsfbox{#1}}

\begin{document}

\draft
\preprint{\vbox{
\hbox{UMD-PP-03-054} }}

  
\title{UNDERSTANDING NEUTRINO MASSES AND MIXINGS IN THE
SEESAW FRAMEWORK\footnote{Based on invited talks at the PASCOS03
conference, Mumbai, India, January, 2003; KITP, Santa Barbara
Neutrino workshop, March, 2003 and Neutrino Telescope conference in
Venice, Italy, March, 2003.}}

\author{R. N. MOHAPATRA}
\address{Department of Physics, University of Maryland, College Park,
MD 20742, USA }
\date{May, 2003}
\maketitle

\begin{abstract}
Understanding the disparate mixing patterns between quarks and leptons is
one of the major challenges in particle theory today. I discuss some of
the ways to understand this difference within the seesaw framework 
using new symmetries of quarks and leptons.  After a brief
introduction to the various types of seesaw formulae, proposals for
understanding large solar and atmospheric mixings for the three mass
hierarchies i.e. normal, inverted and degenerate, are presented and their
implications discussed. 
\end{abstract}

\section{INTRODUCTION}
There is now strong evidence for neutrino oscillations
from the solar and atmospheric observations in Super-Kamiokande,
Homestake, Gallex, SAGE and SNO experiments. The
deficit in the neutrino flux observed in detectors on Earth compared to
theoretical expectations in both the solar and the atmospheric cases 
appear to be well understood if one assumes that
neutrinos produced in the source ($\nu_e$'s in the first case and
$\nu_\mu$'s in the second) oscillate into another species
($\nu_{\mu,\tau}$ in the first case and $\nu_\tau$ in the second) which
are not observable. Laboratory experiments that use accelerator muon neutrinos
as in the K2K experiment and reactor electron anti-neutrinos as
in the Kamland
experiment have also shown deficits in the number of neutrinos compared to
expectations providing not only additional evidence for oscillations 
but also ruling out alternative explanations for the solar and
atmospheric neutrino deficits. 

Thus the phenomenon of neutrino oscillations is now well established. The
questions now are (i) how well do we understand what is observed in the
neutrino experiments and (ii) what does it tell us about the nature of new
physics beyond the standard model ? In this talk I will discuss several
ideas that attempt to answer these questions.

 For inter-species oscillations to
take place, the neutrinos must be massive and must mix
among themselves, with mass differences and mixing
angles that are determined by observations.
 It seems that all the above data (excluding the LSND observations) can be
understood
in terms of oscillations of the three known neutrinos i.e. $\nu_e,\nu_\mu,
\nu_\tau$ among themselves. Since the standard model predicts that
neutrinos are massless,
this evidence for neutrino mass regardless of the details about mixings is
already of fundamental significance and provides one possible direction
for physics beyond the standard model. Most likely possibility is that
evidence for neutrino mass proves the existence of a right handed neutrino
as I elaborate below. The hope is that
by studying the details pattern of neutrino mixings required to understand
observations, we will have have a clearer roadmap of physics beyond the
standard model.

It is worth mentioning that another accelerator experiment, which has also
shown positive evidence for neutrino oscillation i.e. $\bar{\nu}_\mu$ to
$\bar{\nu}_e$ is
the Los Alamos experiment, LSND. This result has not been
confirmed by KARMEN which has also looked for the same process. 
Currently
MiniBoone experiment at Fermilab is searching for this process. If LSND is
confirmed, it will require drastic change in our understanding of
neutrinos- e.g. it will require the existence of sterile neutrinos that
mix with the known neutrinos. We will postpone the discussion of this
evidence in the present talk.

Two other searches for $\nu_e$ oscillations that have yielded negative
results are CHOOZ and PALO-VERDE reactor experiments but they provide
an upper limits on one of the mixing angles (to be
called $U_{e3}$ below) that has
important implications for theories of neutrino masses. Further
experiments are in progress or in planning stages to improve the limits
on $U_{e3}$ and will significantly improve our understanding of the 
theoretical picture.

In this brief overview, I wish to draw attention to some of the
theoretical ideas for understanding neutrino mass and mixing patterns in
extensions of the standard model. This article will focus specifically 
on the seesaw mechanism that seems to provide the simplest way to
understand small neutrino masses\cite{seesaw} and some attempts to
understand large neutrino mixings within the models that rely on the
seesaw mechanism to explain the small neutrino masses.

\subsection{Major theoretical issues in neutrino physics:}
The major issues of interest in neutrino theory are driven by the
following experimental results and conclusions derived from them. 
We will use the notation, where the flavor or weak
eigen states $\nu_{\alpha}$ (with $\alpha~=~e, {\mu}, {\tau}$)
are expressed in terms of the
mass eigenstates $\nu_{i}$ ($i=1, 2, 3)$ as follows: $\nu_{\alpha}
=~\sum_i
U_{\alpha i}\nu_i$. The $U_{\alpha i}$, the elements of the
Pontecorvo-Maki-Nakagawa-Sakata matrix represent the observable mixing
angles in the basis where the charged lepton masses are diagonal.
In any other basis, one has $U =U^{\dagger}_{\ell}U_{\nu}$, where
the matrices $U_{\ell}$ and $U_{\nu}$ are the ones that diagonalize the
charged lepton and neutrino mass matrices respectively.

\subsubsection{Solar neutrinos:}
Thanks to the SNO results on both charged and neutral currents,
and the KAMLAND\cite{kamland} results, there now
appears to be a winner
among the various possible oscillation solutions to the solar neutrino 
puzzle\cite{bahcall}. 
It seems that the so called LMA MSW solution is
preferred over the small angle as well as the low and pure vacuum  
solution\cite{bahcall}. The KAMLAND results\cite{kamland}
 also rules out many of the
nonoscillation as well as the magnetic moment solution to the solar
neutrino problem\cite{valle}. The present range of preferred values of the
oscillation parameters are: $2\times 10^{-5} \leq \Delta
m^2_{\odot}$/eV$^2$ $\leq
4\times 10^{-4}$ and $0.62 \leq sin^22\theta_{\odot}\leq 0.99$ at
3$\sigma$ confidence level. All nonoscillation mechanisms could however be
present at a subdominant level and higher precision experiments are
necessary to test for their presence.

\subsubsection{ Atmospheric neutrinos:}
 Here evidence appears very
convincing that the explanation of observed muon neutrino deficit in
upward going muons as well as the azimuthal angle dependence of this
spectrum involves oscillation of $\nu_{\mu}$
to $\nu_{\tau}$, with $\Delta m^2_{\nu_{\mu}-\nu_{\tau}}\simeq 2.5\times
10^{-3}$ eV$^2$ and maximal mixing $sin^22\theta_A \geq 0.84$ at 99\% c.l.

\subsubsection{ Neutrinoless double beta decay:} Oscillations involve only
mass differences and therefore do not give information on the over all
scale of the neutrino masses. One may hope that neutrinoless double beta
decay may provide this information. It however turns out that this hope
may not be completely justified even if the present limits on lifetime go
up by
two orders of magnitude as is contemplated in many experiments
unless the neutrinos are quasi-degenerate with common mass in the
range bigger than 0.05 eV. 

Nevertheless, neutrinoless double beta decay is an experiment of
fundamental significance since its observation will for the first time
give evidence that neutrino is its own antiparticle and signal the
breakdown of B-L quantum number. Whether a positive signal will lead to 
any conclusion about the detailed pattern of masses is not a simple
question. The point is that in extensions of physics beyond the standard
model, there are several phenomenologically viable mechanisms for
$\beta\beta_{0\nu}$ decay that do
not involve neutrino mass but rather arise from exchanges of heavy
particles such as
doubly charged Higgs bosons, right handed $W$'s or supersymmetric
particles. Once a positive signal is observed, one will have to understand
which contribution has shown up; for this one not only needs 
a precise value of the nuclear matrix element but also some way to isolate
any possible contribution from heavy particle exchange, before any
conclusion regarding the magnitude of the neutrino mass can be deduced.

Searches for $\beta\beta_{0\nu}$ decay has
been going on for several years and a new round of higher precision
experiments are on the verge of being lunched. The most
stringent limits on this decay are from the enriched $^{76}$Ge experiment
by the
Heidelberg-Moscow as well as the IGEX collaborations  and can be
converted to a constraint on masses and
mixing angles as: $\sum_i U^2_{ei} m_i \leq 0.3 $ eV, with an uncertainty
of a factor of 2 to 3 due to nuclear matrix elements. There appears to be
some evidence  for a positive signal in the existing Heidelberg-Moscow
data\cite{klap}, which if confirmed will be a significant
discovery. Presently planned
experiments such as GENIUS, MAJORANA, CUORE, EXO, XMASS and MOON can not
only test this claim but are
expected to push the limit down by one order of magnitude, if they fail
to substantiate the claim.

\subsubsection{ $U_{e3}$:} The CHOOZ and PALO
VERDE reactor experiments mentioned earlier searched for
disappearance of reactor anti-neutrinos. Their null result can be
translated into an upper limit on the $U_{e3}$ parameter i.e. $U_{e3}\leq
0.16-0.2$
for mass differences given by $\Delta m^2\geq 3\times 10^{-3}$ eV$^2$.

All this information can be summarized in the following form for the ${\bf
\bf U}$ matrix (ignoring CP violation):
\begin{eqnarray}
{\bf \bf U}~\simeq~\left(\begin{array}{ccc} c & s & \epsilon \\
-\frac{s+c\epsilon}{\sqrt{2}} & \frac{c-s\epsilon}{\sqrt{2}} &
\frac{1}{\sqrt{2}} \\
\frac{s-c\epsilon}{\sqrt{2}} & \frac{-c-s\epsilon}{\sqrt{2}} &
\frac{1}{\sqrt{2}} \end{array}\right)~.
\end{eqnarray}
where $\epsilon \equiv U_{e3}$.

As far as the mass pattern goes however, there are three possibilities all
equally viable from experimental point of view:

\begin{itemize}

\item (i) normal hierarchy: $m_1\ll m_2 \ll m_3$ ; 

\item (ii) inverted hierarchy : $m_1\simeq -m_2 \gg m_3$ and

\item (iii) approximately degenerate pattern $m_1\simeq m_2 \simeq
m_3$;
\end{itemize}

where $m_i$ are the eigenvalues of the neutrino mass matrix. In
the first case, the atmospheric and the solar neutrino data give direct
information on $m_3$ and $m_2$ respectively. On the other hand,
in the last case, the mass differences between the first and
the second eigenvalues will be chosen to fit the solar neutrino data and
the second and the third which then must be close to each other are given 
the atmospheric neutrino data to be $m_1\simeq m_2 \simeq \sqrt{\Delta
m^2}\simeq 0.05$ eV.

Three of the major theoretical challenges in neutrino physics now are:

\begin{itemize}

\item How does one understand the extreme smallness of the neutrino 
masses ?

\item How does one understand two large mixing angles among
neutrinos given that there is so much similarity between quarks and
leptons at the level of interactions and that the quark mixings are small?

\item What is the mass pattern among the neutrinos and how does one
understand them from a theoretical point of view simultaneously with the
near bimaximal mixng pattern ? In particular, why is $\Delta
m^2_{\cdot}/\Delta m^2_A \ll 1$.

\end{itemize}

\section{SEESAW MECHANISM FOR SMALL NEUTRINO MASSES}
It is well known that in the standard model the neutrino is massless due
to a combination of two reasons: (i) one, its righthanded partner
($\nu_R$) is absent and
(ii) the model has exact global $B-L$ symmetry. Clearly, to understand a
nonzero
neutrino mass, one must give up one of the above assumptions. If one
blindly included a $\nu_R$ to the standard model as a singlet, the status
of neutrino would be parallel to all other fermions in the model and one
would be hard put to understand why its mass is so much smaller than
that of other fermions. Clearly there must be some other new ingredient
that must be added.

A first hint of this new ingredient came from the observation of
Weinberg that if B-L symmetry is broken by some high scale physics, in 
the effective low energy
theory, one can have operators of the form $(LH)^2/M$, where $M$ denotes
the scale of new physics\cite{weinberg}. This after electroweak symmetry
breaking would lead to a neutrino mass $\sim \frac{v^2_{wk}}{M}$. 
The key question now is what is the value of $M$ ?

In the
absence of any B-L violating physics all the way upto the Planck scale
and assuming that nonperturbative Planck scale physics breaks all global
symmetries such as the global B-L symmetry present in the standard model, 
a plausible higher  
dimensional operators takes the form\cite{barbieri} $LHLH/M_{P\ell}$
(where $L$ is a lepton doublet and $H$ is the Higgs doublet). This
afert electrweak symmetry breaking leads to masses for
neutrinos of order $10^{-5}$ eV or less and are therefore not adequate
for understanding observations. Thus a nontrivial extension of the
standard model is called for wherein, the requisite value for $M$
to explain the atmospheric neutrino data ( of
order $10^{15}$ GeV or so) must be the scale of B-L breaking or the
breaking of some other symmetry. A concrete example of a particle that
will provide the ultraviolet completion of the standard model with the
desired neutrino mass operator is to add right handed neutrinos which have
a large Majorana mass. This is the seesaw mechanism\cite{seesaw}, that I
will discuss
below.

One then faces a ``naturalness'' problem
similar to the Higgs mass problem of the standard model i.e. why the
radiative corrections do not send the mass $M$ of the right handed
neutrino to the Planck scale.

An associated question is: is there an indepedent reason for the right
handed neutrino other than the neutrino mass and seesaw mechanism ?
We will see below that there are several candidate symmetries which
are compelling from other arguments and provide a reason for the stability
of the new scale mass $M$ and naturally bring in the right handed
neutrino into the theory. These symmetries are local
symmetries.
\begin{itemize}

\item (i) local $B-L$ and/or

\item (ii) $SU(2)_H$ horizontal symmetry acting on the first two
generations\cite{kuchi};

\item (iii) $SU(3)_H$ horizontal symmetry\cite{kribs}.
\end{itemize}
The most widely discussed example is the local B-L symmetry
but the second case has also very interesting predictions for neutrino
masses. The mass $M$ in these examples is the Majorana mass
of the  right handed neutrinos that break either or both of these
symmetries
(i.e. in the exact symmetry limit the RH neutrinos have zero mass).

\subsection{Quark lepton symmetry and local B-L symmetry}
As the first example of a model with right handed neutrinos ($N_R$),
consider
making the standard model completely quark lepton symmetric by adding one
$N_R$ per generation. This expands the gauge symmetry of the 
electroweak interactions to $SU(2)_L\times U(1)_{I_{3R}}\times U(1)_{B-L}$
or to its full left-right symmetric extension $SU(2)_L\times SU(2)_R
\times U(1)_{B-L}$ symmetry. In the latter case, the fermion doublets
$(u,d)_{L,R}$ and $(\nu, e)_{L,R}$ are assigned
to the left-right gauge group in a parity symmetric manner. The electric
charge formula for the model takes a very interesting form\cite{marshak}:
$Q= I_{3L} + I_{3R} + \frac{B-L}{2}$. It can be concluded from this that
below the scale $v_R$ where the $SU(2)_R\times U(1)_{B-L}$ symmetry
breaks down to the standard model and above the scale of $M_W$, one has
the relation $\Delta I_{3R}~=~-\Delta \frac{B-L}{2}$. This simple
looking relation has the profound consequence
that neutrinos must be  Majorana fermions and that there must be lepton
number violating interactions in nature.
 Furthermore it explains why the right handed neutrino mass is so much
smaller than the Planck mass- it is connected with the breaking of
local $B-L$ symmetry. 

\subsection{Type I vrs type II seesaw}

To see how small neutrino masses are explained, note that
the $\nu_L-\nu_R$ mass matrix for three generations takes the form:
\begin{eqnarray} 
M~=~ \left(\begin{array}{cc} M_{LL} & M_{LR}\\ M^T_{LR} & M_{RR}
\end{array}\right)
\end{eqnarray}
where $M_{RR} = {\bf f}v_R$ is the Majorana mass matrix of the right
handed neutrinos, (${\bf f}$ is the new Yukawa coupling matrix that
determines the right handed neutrino masses). The first term
$M_{LL}~\simeq~{\bf f}\frac{ v^2_{wk}}{v_R}$ is the induced triplet vev
that leads to a direct Majorana
mass matrix for the left handed neutrinos and is characteristic of the
existence of asymptotic parity symmetry. (It would for example be absent
if the local symmetry is  $SU(2)_L\times U(1)_{I_{3R}}\times U(1)_{B-L}$.)
Note that the flavor structure of the induced triplet vev contribution (or
the type II seesaw contribution), is same as for the right handed
neutrino.

 The contribution $M_{LR}\equiv M_D =
{\bf Y}v_{wk}$ is the Dirac mass matrix connecting the left and the right
handed neutrinos. The diagonalization of this mass matrix leads to
following form for the light neutrino masses:
\begin{eqnarray}
M_{\nu}\simeq~{\bf f}\frac{ \lambda v^2_{wk}}{v_R}-\frac{1}{v_R} M^T_D{\bf
f}^{-1}M_D;
\label{seesaw}
\end{eqnarray}
where $M_D$ is defined as ${\cal L}_{mass} = \bar{\nu}_R M_D \nu_L$;
${\bf f}$, the Yukawa coupling matrix that is responsible for the masses
of the heavy right handed neutrinos characterizes the
high scale physics, whereas all other parameters denote physics at the
weak scale. We have called this generalized formula for neutrino masses,
the type II seesaw formula\cite{ma} to distinguish it from the type I
seesaw formula, the 
one that is commonly used in literature where the
first term of Eq. \ref{seesaw} is absent. Important feature
of this formula is that both
terms vanish as $v_R\rightarrow \infty$ and since $v_R \gg v_{wk}$, the
the neutrino masses are much smaller than the charged fermion
masses. As was particularly emphasized in the third paper of
ref.\cite{seesaw}, the
dominance of V-A interaction in the low energy weak processes is now
connected to smallness of neutrino masses.

If in the above seesaw formula, the second term dominates, this leads to
the canonical type I seesaw formula and leads to the often discussed
hierarchical neutrino masses, which in the approximation of small mixings 
lead to $m_{\nu_{i}}\simeq m^2_{f_i}/v_R$, where $f_i$ is either a charged
lepton or a quark depending on the kind of model for neutrinos.

On the other hand, in models where the first term dominates, the neutrino
masses can be almost generation independent unless $f$ itself
has the flavor structure of the charged fermions. For example, if there is
indication for neutrinos being degenerate in mass from
observations, one will have to resort to type II seesaw mechanism for its
understanding, with the first term dominating the neutrino masses.

A further advantage of the right handed neutrino within the B-L symmetry
framework and seesaw mechanism is
that they fit very nicely into grand unified frameworks based on
SO(10) models. The coupling constant unification then provides
a theoretical justification for the high seesaw scale and hence the small 
neutrino masses. Furthermore, the {\bf 16}-dim. spinor representation of
SO(10) has just the right quantum numbers to fit the $\nu_R$ in addition
to the standard model particles of each generation.

\subsection{Double seesaw with a low scale for $B-L$ symmetry}

As we saw from the previous discussion, the conventional seesaw mechanism
requires rather high scale for the B-L symmetry breaking and the
corresponding right handed neutrino mass (of order $ 10^{15}$
GeV). There is however no way at
present to know what the scale of B-L symmetry breaking is. There are
for example models bases on string compactification\cite{langacker} where
the $B-L$ scale is quite possibly in the TeV range. In this case small
neutrino mass can be implemented by a double seesaw mechanism suggested in
Ref.\cite{valle1}. The idea is to take a right handed neutrino $N$ whose
Majorana mass is forbidden by some symmetry and a
singlet neutrino $S$ which has extra quantum numbers which prevent it from
coupling to the left handed neutrino but which is allowed to couple
to the right handed neutrino. One can then write a three by three
neutrino mass matrix in the basis $(\nu, N, S)$ of the form:
\begin{eqnarray}
{ M}~=~\left(\begin{array}{ccc} 0 & m_D & 0 \\ m_D & 0 & M \\ 0 & M
& \mu\end{array}\right)
\end{eqnarray}
For the case $\mu \ll M \approx M_{B-L}$, (where $M_{B-L}$ is the $B-L$
breaking scale), this matrix has one light and two heavy states. The
lightest eigenvalue is given by $m_\nu\sim m_d M^{-1}\mu M^{-1}
m_D$. There is a double suppression by the heavy mass compared to the
usual seesaw mechanism and hence the name double seesaw. 
A generalization of this mechanism to the case of three generations is
straightforward. One important point here is that to keep $\mu\sim m_D$,
one also needs some additional gauge symmetries, which often are a part
of the string models. It can also be used in models with high scale B-L
breaking where the RH neutrino is forbidden by symmetries\cite{raby}

\subsection{$SU(2)_H$ local symmetry and $3\times 2$ seesaw with two
$N_R$'s}

A symmetry among the different generation has often been suspected as a
possible way to understand the different properties of the quarks and
leptons of different
generations. This symmetry for the three generation case could be either
a $U(1)$, $SU(2)$ or an $SU(3)$ local symmetry. Of these
three possibilities, the third one requires that we include additional
fermions to cancel anomalies. Of the remaining two, we choose 
$SU(2)_H$ since it has the following interesting property i.e. 
if it operate on right handed charged
leptons, cancellation of global Witten anomaly requires that we must
introduce
at least two right handed neutrinos $(N_{eR}, N_{\mu R})$ transforming as
a doublet under the group. Thus two right handed neutrinos is the minimal
set required theoretically. Clearly, the mass of the right handed
neutrinos are connected to the breaking of the $SU(2)_H$
symmetry\cite{kuchi}. An additional feature of these matrices is that
they lead to a $3\times 2$ seesaw as compared to the $3\times 3$ seesaw
in the case of the left-right symmetric (or SO(10)) models. We will see
the implications of the $3\times 2$ seesaw later on in this talk. This
could
of course be a part of a B-L like model if $v_H \ll M_{B-L}$.
A distinct feature of the models with 3$\times $2 seesaw is that one of
the light neutrinos is massless. In this sense, in these models all
parameters of a real neutrino mass matrix are determinable by only
oscillation experiments.

\section{MASS MATRIX ANSATZ AND ATTEMPTS TO UNDERSTAND LARGE MIXINGS}

One of the major mysteries of neutrino physics is understanding large
mixing angles needed to explain solar and the atmospheric neutrino
oscillations.
This is because of the simple fact that there is so much similarity in the
interactions between the quarks and leptons and yet quarks mixings between
different generations are of course well known to be very small unlike 
the lepton mixings. 

In the seesaw framework one may attribute the origin of large mixings to
the fact that a central ingredient 
in understanding the neutrino mass matrix is the mass matrix of the right
handed neutrino which reflects high scale physics whereas quark physics is
presumably a low scale physics. From this perspective, one should
``invert'' the seesaw formula and deduce the texture of right handed
neutrino masses from our knowledge of neutrino masses andf mixings 
\cite{falcone}. One then
needs to understand where the relevant right handed neutrino mass
matrix comes from and draw clues from it as to the nature of high scale
physics. An important point is that if the right handed neutrinos
are also responsible for origin of matter via leptogenesis\cite{fuku},
then these conclusions about the RH neutrino mass matrix can in principle
be ``tested'' using this cosmological laboratory.

To proceed further, a good starting point is to search useful mass
matrices for light neutrinos that explain observed mixings.
We first note that in the absence of CP
violation, the symmetric Majorana mass matrix for the light neutrinos
${ M}_\nu$ contains six parameters, whereas observations give only 
five pieces of observation i.e. $\Delta m^2_{A,\cdot}$,
$\theta_{12}\equiv \theta_{\odot}$, $\theta_{23}\equiv \theta_A$ and
$U_{e3}\equiv \theta_{13}$. The absence of the sixth piece of information
is essentially reflected in the fact that the precise mass pattern
(normal, inverted or degenerate) of neutrinos is not known. So to make any
progress, one may try to make ansatzes that reduce the number of
parameters in a mass matrix either (A) by making different elements equal
or (B) putting them to zero in a basis where the charged leptons are
diagonal.

An example of the first strategy is the zeroth order mass matrix
discussed in\cite{nussinov}:
 \begin{eqnarray}
M_{\nu}=\left(\begin{array}{ccc} A+D & F & F \\ F & A & D \\ F & D & A
\end{array}\right)
\end{eqnarray}
This leads to an exact bimaximal pattern with the MNS matrix of the form
\begin{eqnarray}
U_{PMNS}~=~\pmatrix{\frac{1}{\sqrt{2}} & \frac{1}{\sqrt{2}}  & 0\cr
\frac{1}{{2}} &-\frac{1}{{2}} & \frac{1}{\sqrt{2}} \cr \frac{1}{{2}}
&-\frac{1}{{2}} & -\frac{1}{\sqrt{2}} }
\end{eqnarray}
but allows for all different mass
patterns depending on the relative values of the parameters $A, D$ and
$F$.
Since the present data implies that there are deviations from the exact
bimaximal form, this mass matrix must have additional small corrections.

Three different mass patterns can emerge from this mass matrix in
various limits: e.g. (i) for $F \ll A \simeq -D$, one gets the normal
hierarchy; (ii) for $F \gg A, D$, one has the inverted pattern for masses
and (iii) the parameter region $F, D \ll A$ leads to the degenerate case.
An interesting symmetry of this mass matrix is the $\nu_\mu
\leftrightarrow
\nu_\tau$ interchange symmetry, which is obvious from the matrix; but in
the limit where $A=D=0$, there appears a much more interesting symmetry
i.e. the continuous symmetry $L_e-L_\mu-L_\tau$\cite{emutau}. If the
inverted mass matrix is confirmed by future experiments, this symmetry
will provide an important clue to new neutrino related physics beyond the
standard model. Inverted mass pattern is the only case where such an
interesting leptonic symmetry appears.
We explore the implications of this symmetry further in the next section.
But before that, let us explore some other ansatz in thye literature.

Another  mass matrix of interest\cite{gl} has the form
\begin{eqnarray}
M_{\nu}=\left(\begin{array}{ccc} A & F & F \\ F & G & D \\ F & D & G
\end{array}\right)
\end{eqnarray}
which is a four parameter mass matrix which introduces a new parameter
that can be varied to obtain the desired solar neutrino mixing. The
MNS matrix that results from this mass matrix is:
\begin{eqnarray}
U_{PMNS}~=~\pmatrix{c & s  & 0\cr
-\frac{s}{\sqrt{2}} &\frac{c}{\sqrt{2}} & \frac{1}{\sqrt{2}} \cr
-\frac{s}{\sqrt{2}}
&\frac{c}{\sqrt{2}} & -\frac{1}{\sqrt{2}} }
\end{eqnarray}
Clearly as in the case of ref.\cite{nussinov}, if $F \gg A,D,G$, we get
the inverted pattern and for $F,A\ll D,G$, we get the normal mass
hierarchy.
 
One can make the parameters in this mass matrix complex\cite{bmv} as
follows:
\begin{eqnarray}
M_{\nu}=\left(\begin{array}{ccc} A & F & F^* \\ F & G & D \\ F^* & D & G^*
\end{array}\right)
\end{eqnarray}
in which case one gets a $U_{e3}$ imaginary corresponding to maximal
Dirac CP violation, a possibility that has important experimental
implications. There are many other interesting mass matrix
ansatzes\cite{anjan} which have their characteristic predictions.

One lesson that one may learn from these studies is the possible existence
of symmetries, which may shed light on the nature of new physics beyond
the standard model and in all the cases discussed, $L_e-L_\mu-L_\tau$
emerges as a possible candidate as noted earlier. We study the
implications and tests of this symmetry below.

\subsection{Approximate $L_e-L_\mu-L_\tau$ symmetry and neutrino
mixings}
In the exact $L_e-L_\mu-L_\tau$ symmetry limit, the
model not only leads naturally to large solar and atmospheric mixing
angles but
it also leads to vanishing $U_{e3}$ as well as $\Delta m^2_{odot}/\Delta
m^2_A = 0$. Therefore the model raises the hope that a small
$U_{e3}$ as well as the smallness of $\Delta m^2_{odot}/\Delta
m^2_A $ can be understood in a natural manner. One must therefore add
small symmetry breaking terms to this model and examine the consequences.

This question was studied in two papers\cite{bm}. In the second paper of
\cite{bm}, the following
mass matrix for neutrinos was considered that includes
 small $L_e-L_\mu-L_\tau$ violating terms.
\begin{eqnarray}
{ M}_\nu=m~\left(\begin{array}{ccc} z &
c & s\\ c & y & d\\ s & d & x\end{array}\right).
\end{eqnarray}
where $c=cos\theta$ and $s=sin\theta$ and $x,y,z,d \ll 1$ (we allow
$x,y,z,d$ to be random and as large as $0.3$.
The charged lepton mass matrix is chosen to have a diagonal form in this
basis and $L_e-L_\mu-L_\tau$ symmetric.

 In the presence of the small symmetry breaking terms $(x,y,z,d)$,
 we find the following sumrules
involving the neutrino observables and the elements of the neutrino mass
matrix.  The nontrivial ones are:
\begin{eqnarray}
\sin^22\theta_{\odot}~=~1-(\frac{\triangle m_\odot^2}{4\triangle
m_A^2}-z)^2~+~O(\delta^3) \cr
 \frac{\triangle m_\odot^2}{\triangle
m_A^2}~=~2(z+\vec{v}\cdot\vec{x})~+~O(\delta^2)\cr
U_{e3}~=~\vec{A}\cdot(\vec{v}\times\vec{x})~+~O(\delta^3)\cr
\end{eqnarray}
where $\vec{v}=(\cos^2\theta,\sin^2\theta,\sqrt{2}\sin\theta\cos\theta)$,
$\vec{x}=(x,y,\sqrt{2}d)$ and
$\vec{A}~=~\frac{1}{\sqrt{2}}(1,1,0)$. $\delta$
in the preceding equations represents the
small parameters in the mass matrix.

One of the major consequences of these relations is that (i) there is a
close connection between the measured value of the solar mixing angle and
the neutrino mass measured in neutrinoless double beta decay
i.e. $z$; (ii) the
present values for the solar mixing angle can be used to predict the
$m_{\beta\beta}$ for a value of the $\Delta m^2_{\odot}$. For instance,
for $sin^22\theta_{\odot}=0.9$, we would predict
$(\frac{\triangle m_\odot^2}{4\triangle m_A^2}-z) =0.3$. For small $\Delta
m^2_{\odot}$, this implies $m_{\beta\beta}\simeq 0.01$ eV.
The second relation involving the $\Delta m^2_{\odot}/\Delta m^2_A$ in
terms of $x, y , z, d$ tells us that for this to be the case, we must
have strong cancellation between the various small parameters. Given
this, the above $m_{\beta\beta}$ value
is expected to be within the reach of new double beta decay experiments
contemplated. Note however that the $sin^22\theta_{\odot}$
cannot be smaller than $0.9$ in the case of approximate
$L_e-L_\mu-L_\tau$ symmetry.

If the value of  $sin^22\theta_{\odot}$ is ultimately determined to be
less than $0.9$, the question one may ask is whether the idea of
$L_e-L_\mu-L_\tau$ symmetry is dead. The answer is in the negative since
so far we have explored the breaking of $L_e-L_\mu-L_\tau$ symmetry only
in the neutrino mass matrix. It was shown in the first paper of \cite{bm}
that if the
symmetry is broken in the charged lepton mass, one can lower the
$sin^2\theta_{odot}$ to $0.85$ or so.

\subsection{Approximate $L_e-L_\mu-L_\tau$ symmetry from $SU(2)_H$
horizontal symmetry}
It can be shown that an $SU(2)_H$ model for leptons leads quite generally
to an approximate $L_e-L_\mu-L_\tau$ symmetry for neutrinos. As already 
noted, a distinct
feature of $SU(2)_H$ symmetry is that there are two right
handed neutrinos instead of three and therefore
one has a $3\times 2$ seesaw rather than the usual $3\times 3$
one.

To see this in detail, first note that the gauge interactions have the
symmetry $SU(2)_H\times U(1)_{e+\mu+\tau}$ global symmetry. The diagonal 
generator of $SU(2)_H$ is given by $L_e-L_{\mu}$. If we break
horizontal symmetry by an $SU(2)_H$ triplet Higgs $\Delta_H$, then
$L_e-L_\mu$ survives as a gauge symmetry of leptons. We further break the
symmetry by a doublet Higgs $\chi_H$, then the
allowed Yukawa couplings  that contribute to neutrino masses are of the
form $N^c\Delta_H N^c$, $L_\tau H_u\chi_HN^c$. Note that these two terms
reduce the above global symmetry to $SU(2)_H\times U(1)_{\tau}$. The vevs
of these Higgs fields i.e. $<\Delta_H,3>\neq 0$ and $\chi_{H,2}\neq 0$
reduces this symmetry down to $L_e-L_\mu -L_\tau$. This is the major
reason why this model leads to an inverted hierarchy and also two large
mixings in zeroth order as desired. Thus if experiments confirm the
inverted hierarchy and a possible $L_e-L_\mu-L_\tau$ symmetry for leptons,
it may be signal for the local $SU(2)_H$ symmetry at a high scale.

 The charged lepton
masses arise from the couplings of the form $LH_d\chi_H\tau^c$ and $L_\tau
H_d \chi_H E^c$. The second term breaks $L_e-L_\mu-L_\tau$ symmetry and is
responsible departure from exact maximal mixing angle in the 12 sector as
well can contribute to solar mass splittings.

Using the discussion of the above paragraph, the Dirac mass
of the neutrino as well as the righthanded neutrino mass matrix can be
seen to lead\cite{kuchi} to 5$\times $5 mass matrix for heavy and light
neutrinos of the form:
\begin{eqnarray}
M_{\nu_L,\nu_R}~=~\left(\begin{array}{ccccc} 0 & 0 & 0 & h_0\kappa_0 & 0\\
0 & 0 & 0 & 0 & h_0\kappa_0\\ 0 & 0 & 0 & h_1\kappa_1 & h_1 \kappa_2 \\
h_0\kappa_0 & 0 & h_1\kappa_1 & 0 & fv'_H \\ 0 & h_0\kappa_0 & h_1\kappa_2
& fv'_H & 0 \end{array} \right)
\end{eqnarray}
After seesaw diagonalization, it leads to the light neutrino mass matrix
of the form:
\begin{eqnarray}
{ M}_{\nu}~=~-M_D M^{-1}_R M^T_D
\end{eqnarray}
where $M_D~=~\left(\begin{array}{cc} h_0\kappa_0 & 0 \\ 0 & h_0\kappa_0\\
h_1\kappa_1 & h_1\kappa_2
\end{array}\right)$; $M^{-1}_R~=~\frac{1}{fv'_H}\left(\begin{array}{cc} 0
&
1\\1 & 0 \end{array}\right)$. The resulting light Majorana neutrino mass
matrix ${ M}_{\nu}$ is given by:
\begin{eqnarray}
{ M}_{\nu}~=~-\frac{1}{fv'_H}\left(\begin{array}{ccc} 0 &
(h_0\kappa_0)^2 & h_0h_1\kappa_0\kappa_2\\ (h_0\kappa_0)^2 & 0 &
h_0h_1\kappa_0\kappa_1 \\ h_0h_1\kappa_0\kappa_2 & h_0h_1 \kappa_0\kappa_1
& 2h^2_1\kappa_1\kappa_2 \end{array}\right)
\end{eqnarray}
First of all as discussed before, this leads to one neutrino which is
massless. Also note that as $\kappa_1\rightarrow 0$, the mass matrix
acquires $L_e-L_\mu-L_\tau$ symmetry and of course has $\Delta
m^2_\odot=0$. The smallness of  $\Delta m^2_\odot$ thus implies that
$\kappa_1\ll \kappa_{0,2}$. Also in the limit the solar mixing angle
is $\pi/4$. 

To get the physical neutrino mixings, we also need the charged lepton mass
matrix defined by $\bar{\psi}_L { M}_\ell \psi_R$. There are two
possibilities for $M_{\ell}$ in our model:

\noindent {\it Case (i)}:

\begin{eqnarray}
{ M}_{\ell}~=~\left(\begin{array}{ccc} h'_2\kappa_0 & 0 &-h'_1\kappa_2
\\ 0 & h'_2\kappa_0 & h'_1\kappa_1 \\ h'_4\kappa_1 & h'_4\kappa_2 &
h'_3\kappa_0 \end{array}\right)
\end{eqnarray}

\noindent {\it Case (ii)}

\begin{eqnarray}
{ M}_{\ell}~=~\left(\begin{array}{ccc} h'_2\kappa_0 & 0 &h'_1\kappa_1
\\ 0 & h'_2\kappa_0 & h'_1\kappa_2 \\ -h'_4\kappa_2 & h'_4\kappa_1 &
h'_3\kappa_0 \end{array}\right)
\end{eqnarray}
There are contributions to neutrino mixings
coming from the charged lep[ton sector, which will help us to get a lower
value for $sin^22\theta_\odot$. Generically, we get on including the
charged lepton contributions a correlation between the solar mixing angle
and $U_{e3}$ as follows: $\theta_{\odot}\approx
\frac{\pi}{4}-U_{e3}$\cite{seidl}.

\section{LARGE MIXINGS IN MODELS WITH QUARK-LEPTON UNIFICATION}
The $L_e-L_\mu-L_\tau$ model discussed above treats the quarks and leptons
on a fundamentally different footing. On the other hand it could be that
at very short distances there is quark lepton unification\cite{pati}. I
give below two of a number of ideas, where models with quark lepton
symmetry can lead to large neutrino mixings. In the models discussed below
large mixings arise dynamically and without need for any extra symmetries. 
 
\subsection{ Radiative magnification of mixing angles}
In this class of models dynamics of radiative corrections plays an
essential role in
understanding the maximal mixings. The basic idea is that at the seesaw
scale, all mixings angles are small, a situation quite natural if the
pattern of ${\bf f}$ Yukawa coupling is similar to the quark sector. 
 Since the observed neutrino mixings are weak scale
observables, one must extrapolate\cite{babu1} the seesaw scale mass
matrices to the weak scale and recalculate the mixing angles. 

The extrapolation formula is 
\begin{eqnarray}
{ M}_{\nu}(M_Z)~=~ {\bf I}{ M_{\nu}} (v_R) {\bf I}\\
where~~~~~~~~{\bf I}_{\alpha \alpha}~=~
\left(1-\frac{h^2_{\alpha}}{16\pi^2}\right)
\end{eqnarray}
Note that since $h_{\alpha}= \sqrt{2}m_{\alpha}/v_{wk}$ ($\alpha$ being
the charged lepton index), in the extrapolation only the $\tau$-lepton
makes a difference. In the MSSM, this increases the ${ M}_{\tau\tau}$
entry of the neutrino mass matrix and essentially leaves the others
unchanged. It was shown
in ref.\cite{balaji} that if the muon and the tau neutrinos are
nearly degenerate in mass at the seesaw scale, in supersymmetric
theories, the $tan\beta\geq
5$, the radiative corrections can become large enough so that at the weak
scale the two diagonal elements of ${ M}_{\nu}$ which were nearly
equal but different at the seesaw
scale become much more degenerate. This leads to an enhancement of the
mixing angle to become almost maximal and a
solution to the atmospheric neutrino deficit emerges
even though at the seesaw scale, the mixing angles were small.

This can also be seen from the renormalization group equations when they
are written in the mass basis\cite{casas}. Denoting the mixing angles as
$\theta_{ij}$ where $i,j$ stand for generations, the equations are:
\noindent
\begin{eqnarray}
\dsodt&=&-F_{\tau}{c_{23}}^2\left(
-s_{12}U_{\tau1}D_{31}+c_{12}U_{\tau2}D_{32}
\right),\label{eq3}\\
\dstdt&=&-F_{\tau}c_{23}{c_{13}}^2\left(
c_{12}U_{\tau1}D_{31}+s_{12}U_{\tau2}D_{32}
\right),\label{eq4}\\
\dsthdt&=&-F_{\tau}c_{12}\left(c_{23}s_{13}s_{12}U_{\tau1}
D_{31}-c_{23}s_{13}c_{12}U_{\tau2}D_{32}\right.\nonumber \\
&&\left.+U_{\tau1}U_{\tau2}D_{21}\right).\label{eq5}\end{eqnarray}
\noindent
where $D_{ij}={\left(m_i+m_j)\right)/\left(m_i-m_j\right)}$ and
$U_{\tau 1,2,3}$ are functions of the neutrino mixings angles. The
presence
of $(m_i-m_j)$ in the denominator makes it clear that as $m_i\simeq m_j$,
that particular coefficient becomes large and as we extrapolate from the
GUT scale to the weak scale, small mixing angles at GUT scale become large
at the weak scale. 

Furthermore, this happens only if the experimental observable $\Delta
m^2_{23}\leq 0$
a possibility can be tested in contemplated long base line experiments.
Also for this mechanism to work, the overall scale of neutrino masses must
be in the range of 0.1 eV or so making the idea testable in forthcoming
double beta decay experiments.

Several comment are in order: (i) to get a near degenerate mass spectrum
without additional assumtions, one must use the type II seesaw mechanism
as in Eq. (3); (ii) An interesting question is whether this mechanism can
be extended
to the case of three generations and whether it can explain the
bimaximal pattern also. It has been shown recently that indeed this can
work for three generations\cite{par}, where identifying the seesaw scale
neutrino mixing angles with the corresponding quark mixings and assuming
quasi-degenerate neutrinos, it is found\cite{par} that weak scale solar
and atmospheric angles get magnified to the desired level while due to the
extreme smallness of $V_{ub}$, the magnified value of $U_{e3}$ remains
within its present upper limit. In figure 1, we show the evolution of the
mixing angles to the weak scale.
 \begin{figure} 
\epsfxsize=8.5cm
\epsfbox{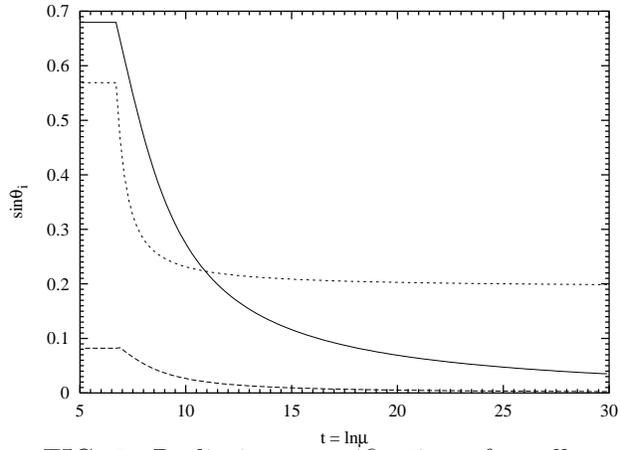}
\caption{Radiative magnification of small quark-like neutrino mixings at
the see-saw scale to bilarge values at low energies. The solid, dashed and
dotted lines represent
$\sin\theta_{23}$, $\sin\theta_{13}$, and $\sin\theta_{12}$,
respectively.}
\label{fig1}
\end{figure}

  A second
recent work\cite{gautam} has used the techniques of ref. \cite{balaji}
to study radiative magnification of solar angle in texture zero
neutrino mass matrices. In this example, the atmospheric neutrino mixing
is an input but solar angle is dynamically magnified.

\subsection{A minimal SO(10) model}

Another suggestion for understanding large atmospheric mixing has been
made within a class of SO(10) models, which are strongly suggested by
 local B-L symmetry, large seesaw scale and grand unification ideas.
The basic ingredients of this suggestion are the following properties 
of the SO(10) model: (i) that one can construct a minimal SO(10) model
with only two multiplets that couple to fermions i.e. {\bf 10} and {\bf
126} and another that breaks SO(10) down to the left-right model. The
second breaks the B-L symmetry and the first the electroweak
symmetry. (ii) A second property of SO(10) models \cite{babu} is that {\bf
126} contains submultiplets that not only contribute to
charged fermion but also to the left and right handed Majorana
masses ($M_{LL},M_{RR}$ respectively in Eq. (2)) for the neutrinos. This
leads to a
trmendous reduction of  the number of arbitrary parameters in the model,
as we will see below.

There are only two Yukawa coupling matrices in this model: (i) $h$ for
the {\bf 10} Higgs and (ii) $f$ for the {\bf 126} Higgs. 
SO(10) has the property that the Yukawa couplings involving the {\bf 10}
and {\bf 126} Higgs representations are symmetric. Therefore
if we ignore CP violation and work in a basis where one of these two sets
of Yukawa coupling matrices is diagonal, then it will have
only nine parameters. Noting the fact that the (2,2,15) submultiplet of
{\bf 126} has a standard model doublet that contributes to charged fermion
masses, one can write the quark and lepton mass matrices as
follows\cite{babu}:
\begin{eqnarray}
M_u~=~ h \kappa_u + f v_u \\  \nonumber
M_d~=~ h \kappa_d + f v_d \\  \nonumber
M_\ell~=~ h \kappa_d -3 f v_d \\  \nonumber
M_{\nu_D}~=~ h \kappa_u -3 f v_u \\
\end{eqnarray}
where $\kappa_{u,d}$ are the vev's of the up and down Higgs vevs of the
standard model doublets in {\bf 10} Higgs and $v_{u,d}$ are the
corresponding vevs for the same doublets in {\bf 126}.
Note that there are 13 parameters in the above equations and there are 13
inputs (six quark masses, three lepton masses and three quark mixing
angles and weak scale). Thus all parameters of the model that go into
fermion masses are determined.

 To determine the light neutrino masses, we use the seesaw
formula in Eq. (3), where the {\bf f} is nothing but the {\bf 126}
Yukawa coupling. Thus all parameters that give neutrino mixings except an
overall scale are determined. These models were extensively discussed in
the last decade\cite{last}. Initially CP phases were ignored and more
recently CP phases have been included in the analysis.

A very interesting point regarding these models has been noted
in Ref.\cite{bajc}, where it is pointed out that if the
direct triplet term in type II seesaw dominates, then it provides a very
natural understanding of the large atmospheric mixing angle without
invoking any symmetries. A simple
way to see this is to note that when the triplet term dominates the seesaw
formula,  we have the neutrino mass matrix ${ M}_\nu \propto f$,
where $f$ matrix is the {\bf 126} coupling to fermions discussed earlier.
Using the above equations, one can derive the following
sumrule (sumrule was already noted in the second reference of
\cite{last}):
\begin{eqnarray}
{ M}_\nu~=~ c (M_d - M_\ell)
\label{key}
\end{eqnarray}
Now quark lepton symmetry implies that for the second and third
generation, the $M_{d,\ell}$ have the following general form:
\begin{eqnarray}
M_d~=~\left(\begin{array}{cc}\epsilon_1 & \epsilon_2\cr \epsilon_2 & m_b
\end{array}\right)
\end{eqnarray}
and
\begin{eqnarray}
M_\ell~=~\left(\begin{array}{cc}\epsilon'_1 & \epsilon'_2\cr \epsilon'_2
& m_\tau\end{array}\right)
\end{eqnarray}
where $\epsilon_i \ll m_{b,\tau}$ as is required by low energy
observations. It is well known that in supersymmetric theories, when low
energy quark and lepton masses are extrapolated to the GUT scale, one gets
approximately that $m_b\simeq m_\tau$. One then sees from the above
sumrule for neutrino masses that all entries for the neutrino mass matrix
are of the same order leading very naturally to the atmospheric mixing
angle to be large. Thus one has a natural understanding of the large
atmospheric neutrino mixing angle. No extra symmetries are assumed for
this purpose. 

For this model to be a viable one for three generations, one
must show that the minimal SO(10) model with triplet vev dominated seesaw
formula indeed can give a large $\theta_{12}$ and a small
$\theta_{13}$. This has been shown in a recent paper\cite{goh}. It was
shown that this is indeed the case. To see roughly how this comes about,
let us work in the basis where the down quark mass matrix is diagonal. All
the quark mixing effects are then in the up quark mass matrix i.e.
$M_u~=~U^T_{CKM}M^d_u U_{CKM}$. Note further that the minimality of the
Higgs content leads to the following sumrule among the mass matrices:
\begin{eqnarray}
k \tilde{M}_{\ell}~=~r\tilde{ M}_d +\tilde{ M}_u
\end{eqnarray}
where the tilde denotes the fact that we have made the mass matrices
dimensionless
by dividing them by the heaviest mass of the species i.e. up quark mass
matrix by $m_t$, down quark mass matrix by $m_b$ etc. $k,r$ are functions
of the symmetry breaking
parameters of the model. 
Using the Wolfenstein parameterization for quark mixings, we can conclude
that that we have
\begin{eqnarray}
M_{d,\ell}~\approx ~m_{b,\tau}\pmatrix{\lambda^3 & \lambda^3
&\lambda^3\cr \lambda^3 & \lambda^2& \lambda^2 \cr \lambda^3 & \lambda^2 &
1}
\end{eqnarray}
where $\lambda \sim 0.22$ and the matrix elements are supposed to give
only the approximate order of magnitude. As we extrapolate the quark
masses to the GUT scale, due to the fact that $m_b-m_\tau \approx
m_{\tau}\lambda^2$ for some value of tan$\beta$, the neutrino mass matrix
$M_\nu~=c(M_d-M_\ell)$ takes roughly the form
\begin{eqnarray}
M_{\nu}~=c(M_d-M_\ell)\approx ~m_0\pmatrix{\lambda^3 & \lambda^3
&\lambda^3\cr \lambda^3 & \lambda^2 & \lambda^2 \cr \lambda^3 & \lambda^2
& \lambda^2}
\end{eqnarray}
 It is then easy to see from this mass matrix that both the $\theta_{12}$
(solar angle) and $\theta_{23}$ (the atmospheric angle) are now large. The
detailed magnitudes of these angles of course depend on the details of the
quark masses at the GUT scale. Using the extrapolated values of the quark
masses and mixing angles to the GUT scale, the
predictions of this model for various oscillation parameters are given in
Fig. 1,2 and 3 in a self expalanatory notation. Note specifically the
prediction in Fig. 4 for $U_{e3}$ which can be tested in MINOS as well as
other planned Long Base Line neutrino experiments such as Numi-Off-Axis,
JHF etc. There is a simple explanation of why the $U_{e3}$ comes out to be
large. This can also be seen from the mass sumrule in
Eq.\ref{key}. Roughly, for a matrix with hierarchical eigen values as is
the case here, the mixing angle $tan 2\theta_{13}\sim
\frac{M_{\nu,13}}{M_{\nu,33}}\simeq \frac{\lambda^3
m_\tau}{m_b(M_U)-m_\tau(M_U)}$. Since to get large mixings, we need
$m_b(M_U)-m_\tau(M_U)\simeq m_\tau \lambda^2$, we see that $U_{e3}\simeq
\lambda$ upto a factor of order one. Indeed the detailed calculations lead
to $0.16$ which is not far from this value.

\begin{figure}
\begin{center}
\epsfxsize8cm\epsffile{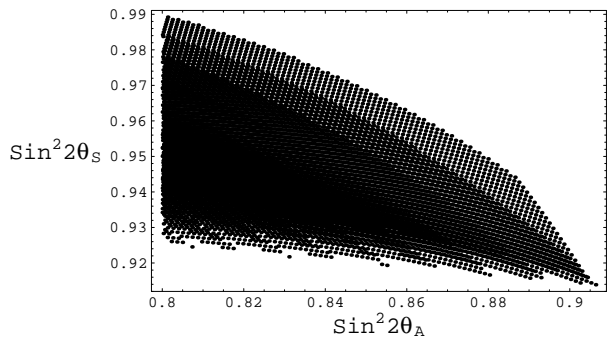}
\caption{
The figure shows the predictions of the minimal SO(10) model for
$sin^22\theta_{\odot}$ and
$sin^22\theta_A$ for the presently range of quark masses. Note that
$sin^22\theta_{\odot}\geq 0.9$ and $sin^22\theta_A\leq 0.9$
\label{fig:cstr1}}
\end{center}
\end{figure}

\begin{figure}
\begin{center}
\epsfxsize8cm\epsffile{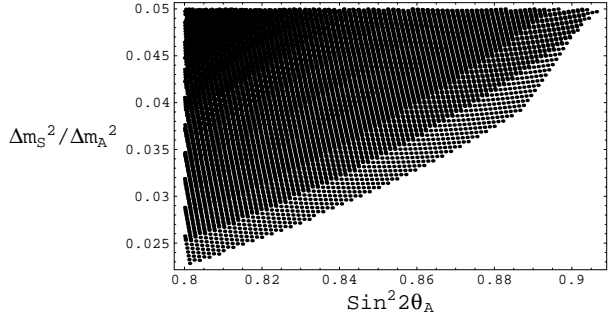}
\caption{
The figure shows the predictions of the minimal SO(10) model for
$sin^22\theta_{A}$ and
 $\Delta m^2_{\odot}/\Delta m^2_{A}$ for the range of quark masses and
mixings that fit charged lepton masses.
\label{fig:cstr2}}
\end{center}
\end{figure}

\begin{figure}
\begin{center}
\epsfxsize8cm\epsffile{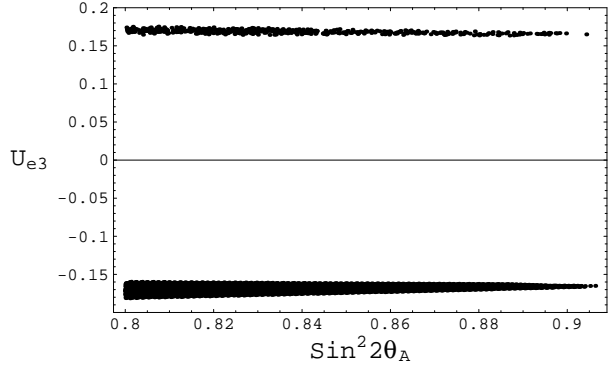}
\caption{
The figure shows the predictions of the minimal SO(10) model for
$sin^22\theta_{A}$
and $U_{e3}$ for the allowed range of parameters in the model. Note that
$U_{e3}$ is very close to the upper limit allowed by the existing reactor
experiments.
\label{fig:cstr3}}
\end{center}
\end{figure}

So far in this work, we assumed that all Yukawa couplings are real. Our
procedure can be easily generalized to the case when they are all complex.
The range of predictions for solar and atmospheric mixing angles become
larger although the prediction for $U_{e3}$ remains around 0.14 to 0.16.

\section{CONCLUSION}
In conclusion, the seesaw mechanism appears by far to be the
simplest way to understand the small neutrino masses. The large right
handed neutrino mass implied by this also helps in understanding origin of
matter in the universe. Our understanding of mixings on the other hand, is
at a very preliminary level. A
particular challenge to theorists is to understand the so called bimaximal
mixing pattern, which is emerging as the favorite. Several symmetry and
dynamical approaches to understanding large mixings are noted. Also a
minimal SO(10) model whose predictions are currently in accord but
testable in
near future is also presented. On the experimental side, high precision
search for $U_{e3}$ and neutrinoless double beta decay will provide
important ways to distinguish between the different models.

This work has been supported by the National Science Foundation grant no.
PHY-0099544. I would like to thank K. S. Babu, B. Dutta, H. S. Goh,
R. Kuchimanchi, S.-P. Ng,
M. K. Parida, G. Rajasekaran for collaboration and discussions and
G. Senjanovi\'c for discussions. I am also grateful to Milla Baldo Ceolin
for kind hospitality at the Neutrino Telescope meeting in Venice.

\end{document}